\documentclass[12pt,twoside]{article}

\usepackage{amssymb}
\usepackage{amsmath}
\usepackage{wrapfig}
\usepackage{graphicx}
\usepackage[eqsecnum]{cmp2e}
\usepackage{citesort}

\title[GF Formalism for Highly Correlated Systems]%
{Green's Function Formalism for \protect\\
Highly Correlated Systems}

\author{F. Mancini and A. Avella}
\address{Dipartimento di Fisica ``E.R. Caianiello'' - Unit\`{a}
CNISM di Salerno \\
Universit\`{a} degli Studi di Salerno, I-84081 Baronissi (SA),
Italy}

\begin{document}

\maketitle

\begin{abstract}
We present the Composite Operator Method (COM) as a modern approach
to the study of strongly correlated electronic systems, based on the
equation of motion and Green's function method. COM uses propagators
of composite operators as building blocks at the basis of
approximate calculations and algebra constrains to fix the
representation of Green's functions in order to maintain the
algebraic and symmetry properties.
\keywords Strongly Correlated Systems, Green's Function Formalism,
Equations of Motion Approach, Composite Operator Method
\pacs 71.10.-w; 71.27.+a; 71.10.Fd
\end{abstract}

\section{Introduction}

The Green's function method is a very convenient formalism in
condensed matter physics, and many progresses have been achieved in
the last fifty years. When applied to interacting systems, such an
approach is usually based on the hypothesis that the interaction
among the particles is weak and can be treated in the framework of
some perturbation schemes. In this line of thinking a consolidated
scheme has been constructed, mostly based on diagrammatic
expansions, Wick's theorem, Dyson equation, and so on. However, in
the last few decades new materials with unconventional properties
have been discovered. It is believed that the origin of such
anomalous behaviors is generally due to strong electronic
correlations in narrow conduction bands \cite{Anderson_87}. In this
line of thinking many analytical methods have been developed for the
study of strongly correlated electron systems \cite{Fulde_95}. The
main difficulties are connected with the absence of any obvious
small parameter in the strong coupling regime and with the
simultaneous presence of itinerant and atomic aspects. The concept
that breaks down is the existence of the electrons as particles with
some well-defined and intrinsic properties. The presence of
interaction modifies the properties of the particles: what are
observed are new particles with new peculiar properties entirely
determined by the dynamics and by the boundary conditions. These new
objects appear as the final result of the modifications imposed by
the interactions on the original particles and contain, by the very
beginning, the effects of correlations. The choice of new
fundamental particles, whose properties have to be self-consistently
determined by dynamics, symmetries and boundary conditions, becomes
relevant.

As a simple example, let us consider an atomic system described by
the Hamiltonian
\begin{equation}\label{1.1}
H=-\mu \sum_\sigma   \varphi _\sigma ^\dag \varphi _\sigma +V\varphi
_\uparrow ^\dag \varphi _\downarrow ^\dag \varphi _\downarrow
\varphi _\uparrow
\end{equation}
$\varphi _\sigma$ denotes an Heisenberg electronic field with spin
$\sigma =\uparrow,\,\downarrow$, satisfying canonical
anticommutation relations; $\mu $ is the chemical potential and $V$
is the strength of the interaction. This model is exactly solvable
in terms of the operators
\begin{equation}\label{1.2}
\xi _\sigma =\varphi _\sigma \varphi _{-\sigma }\varphi _{-\sigma
}^\dag \quad \quad \quad \eta _\sigma =\varphi _\sigma \varphi
_{-\sigma }^\dag \varphi _{-\sigma }
\end{equation}
which are eigenoperators of the Hamiltonian
\begin{equation}\label{1.3}
\mathrm{i} {\partial \over {\partial t}}\xi =[\xi ,H]=-\mu \xi \quad
\quad \quad \mathrm{i}{\partial  \over {\partial t}}\eta =[\eta
,H]=-(\mu -V)\eta
\end{equation}

Due to the presence of the interaction, the original electrons
$\varphi _\sigma$ are no more observables and new stable elementary
excitations, described by the field operators $\xi $ and $\eta $,
appear. Due to the $V$-interaction, two sharp features develop in
the band structure: the energy level $E=-\mu$ of the bare electron
splits in the two levels $E_1=-\mu $ and $E_2=V-\mu $. The bare
electron reveals itself to be precisely the wrong place to start. A
perturbative solution will never give the band splitting.

On the basis of this evidence one can be induced to move the
attention from the original fields to the new fields generated by
the interaction. The operators describing these excitations, once
they have been found, can be written in terms of the original ones
and are known as composite operators.

The convenience of developing a formulation to treat composite
excitations as fundamental objects has been noticed for the
many-body problem of condensed matter physics since long time.
Recent years have seen remarkable developments in many-body theory
in the form of an assortment of techniques that may be termed
composite particle methods. The beginnings of these types of
techniques may be traced back to the work of Bogolubov
\cite{Bogoliubov_47} and later to that of Dancoff \cite{Dancoff_50}.
The work of Zwanzig \cite{Zwanzig_61}, Mori \cite{Mori_65} and
Umezawa \cite{Umezawa_93} has to be mentioned. Closely related to
this work is that of Hubbard
\cite{Hubbard_63,Hubbard_64,Hubbard_64a}, Rowe \cite{Rowe_68}, Roth
\cite{Roth_69} and Tserkovnikov
\cite{Tserkovnikov_81,Tserkovnikov_81a}. The slave boson method
\cite{Barnes_76,Coleman_84,Kotliar_86}, the spectral density
approach \cite{Kalashnikov_69,Nolting_72} and the composite operator
method (COM)
\cite{Ishihara_94,Mancini_94,Mancini_95,Mancini_95a,Mancini_95b,Avella_98,Mancini_00,Mancini_04}
are also along similar lines. This large class of theories is
founded on the conviction that an analysis in terms of elementary
fields might be inadequate for a system dominated by strong
interactions.

All these approaches are very promising because all the different
approximation schemes are constructed on the basis of interacting
particles: some amount of the interaction is already present in the
chosen basis and permits to overcome the problem of finding an
appropriate expansion parameter. However, one price must be paid. In
general, the composite fields are neither Fermi nor Bose operators,
since they do not satisfy canonical (anti)commutation relations, and
their properties, because of the inherent definition, must be
self-consistently determined. They can only be recognized as
fermionic or bosonic operators according to the number, odd or even,
of the constituting original electronic fields. New techniques of
calculus have to be developed in order to treat with composite
fields. In developing perturbation calculations, where the building
blocks are now the propagators of composite fields, the consolidated
scheme (diagrammatic expansions, Wick's theorem, \ldots) gives rise
to very complicated approaches \cite{Izyumov_90} whose application
is far from being easy. The formulation of the Green's function
method must be revisited and new frameworks of calculations have to
be formulated.

\section{Green's Function and Equation of Motion Formalism}

Let us consider a system of $N_e$ interacting Wannier-electrons
residing on a Bravais lattice of $N$ sites, spanned by the vectors
$R_\mathbf{i}=\mathbf{i}$. We ignore the presence of magnetic
impurities and restrict the analysis to single-band electron models.
The generalization of the formalism to more complex systems is
straightforward [see for example
\cite{Matsumoto_94,Mancini_00e,Villani_00,Avella_02,Bak_02}]. The
system is enclosed in a finite but macroscopically large volume V
and is supposed to be in a thermodynamical equilibrium state at a
finite temperature T. In a second quantization scheme this system is
described by a certain Hamiltonian
\begin{equation}\label{2.1}
H=H[\varphi (i)]
\end{equation}
describing, in complete generality, the free propagation of the
electrons and all the interactions among them and with external
fields. $\varphi (i)$ denotes an Heisenberg electronic field [
$i=(\mathbf{i},t)$] satisfying canonical anticommutation relations
derived from the Pauli principle
\begin{equation}\label{2.2}
\{ \varphi _\sigma (\mathbf{i},t),\varphi _{\sigma }^\dag
(\mathbf{j},t)\} =\delta _{\mathbf{ij}}\delta _{\sigma \sigma }\quad
\quad \quad \{ \varphi _\sigma (\mathbf{i},t),\varphi _{\sigma
}(\mathbf{j},t)\} =0
\end{equation}

Any physical property of the system can be connected to the
expectation value of a specific operator. The expectation value of
an arbitrary operator $A=A[\varphi(i)]$ can be computed, for the
grand canonical ensemble, by means of
\begin{equation}\label{2.3}
\langle A \rangle = {{Tr[e^{-\beta (H-\mu \hat N)}A]} \over
{Tr[e^{-\beta (H-\mu \hat N)}]}}
\end{equation}
where the trace implies a sum over a complete set of states in the
Hilbert space. $\hat N=\sum_{i,\sigma }  \varphi _\sigma ^\dag
(i)\varphi _\sigma (i)$ is the total number operator, $\beta
=(k_BT)^{-1}$ is the inverse temperature, $\mu $ is the chemical
potential which is fixed in order to give the chosen average number
of particles $N_e= \langle \hat N \rangle$.

To evaluate the expectation value $\langle A\rangle $, it is
possible to use the equation of motion
\begin{equation}\label{2.4}
\mathrm{i}{\partial  \over {\partial t}}\varphi (i)=[\varphi (i),H]
\end{equation}
in order to derive one or more equations for this quantity or,
better, for the corresponding Green's functions, as explained below.
However, the equation of motion (\ref{2.4}) generates higher-order
operators and more and more complex equations are needed. The
traditional approximation schemes, often based on perturbative
calculations, use as building blocks the noninteracting propagators.
The mean-field formulation, which corresponds to a linearization of
the equation of motion (\ref{2.4}), also belongs to this category.

On the hypothesis that the original fields are not a good basis, we
choose a set of composite fields $\{ \psi (i)\}$ in terms of which a
perturbation scheme will be constructed. Firstly, we choose the set
$\psi (i)$ according to the physical properties we want to study.
Roughly, the properties of electronic systems can be classified in
two large classes: single particle properties, described in terms of
fermionic propagators, and response functions, described in terms of
bosonic propagators. These two sectors, fermionic and bosonic, are
not independent but interplay each other, and a fully
self-consistent solution usually requires that both sectors are
simultaneously solved. Once the sector, fermionic or bosonic, has
been fixed, we have several criteria for the choice of the new
basis. In constructing the composite fields no recipe can be given
without thinking to its drawbacks, but many recipes can assure a
correct and controlled description of relevant aspects of the
dynamics. One can choose: the higher order fields emerging from the
equations of motion (i.e., the conservation of some spectral moments
is assured), the eigenoperators of some relevant interacting terms
(i.e., the relevant interactions are treated exactly), the
eigenoperators of the problem reduced to a small cluster, \ldots

Let $\psi (i)$ be a $n$-component field
\begin{equation}\label{2.5}
\psi (i)= \left( \begin{array}{c}
\psi _1(i) \\
\vdots     \\
\psi _n(i)
\end{array} \right)
\end{equation}
We do not specify the nature, fermionic or bosonic, of the set $\{
\psi (i)\}$. In the case of fermionic operators it is intended that
we use the spinorial representation
\begin{equation}\label{2.6}
\psi _m(i)=\left(\begin{array}{c}
\psi _{\uparrow m}(i) \\
\psi _{\downarrow m}(i)
\end{array} \right)
\quad \quad \quad \psi _m^\dag (i)=\left(\psi _{\uparrow m}^\dag
(i),\,\psi _{\downarrow m}^\dag (i)\right)
\end{equation}

The dynamics of these operators is governed by the given Hamiltonian
$H=H[\varphi (i)]$ and can be written as
\begin{equation}\label{2.7}
\mathrm{i}{\partial  \over {\partial t}}\psi (i)=[\psi (i),H]=J(i)
\end{equation}

It is always possible to decompose the source $J(i)$ under the form
\begin{equation}\label{2.8}
J(i)=\varepsilon (-i\nabla )\psi (i)+\delta J(i)
\end{equation}
where the linear term represents the projection of the source on the
basis $\{ \psi \}$ and is calculated by means of the equation
\begin{equation}\label{2.9}
\langle [\delta J(\mathbf{i},t),\psi ^\dag (\mathbf{j},t)]_\eta
\rangle =0
\end{equation}
Here $\eta =\pm 1$; usually, it is convenient to take $\eta =1$
($\eta =-1$ ) for a  fermionic (bosonic) set $\psi (i)$ (i.e., for a
composite field constituted of an odd (even) number of original
fields) in order to exploit the canonical anticommutation relations
of $\{ \psi (i)\}$; but, in principle, both choices are possible.
Accordingly, we define
\begin{equation}\label{2.10}
\left[ A , B \right]_\eta = \left \{
\begin{array}{l}
\left\{ A , B \right\} =AB+BA \quad for \quad \eta =1 \\
\left[ A , B \right] = AB-BA \quad for \quad \eta =-1
\end{array} \right.
\end{equation}
$\langle \cdots \rangle $ denotes the quantum statistical average
over the grand canonical ensemble, according to Eq.~(\ref{2.3}).
Hereafter, unless otherwise specified, time and space translational
invariance will be considered. The action of the derivative operator
$\varepsilon (-\mathrm{i}\nabla )$ on $\psi (i)$ is defined in
momentum space
\begin{equation}\label{2.11}
\varepsilon (-\mathrm{i}\nabla )\psi (i)=\varepsilon
(-\mathrm{i}\nabla ){1 \over {\sqrt N}}\sum_\mathbf{k}
 e^{\mathrm{i} \mathbf{k} \cdot \mathbf{R_i}}\psi (\mathbf{k},t)={1
\over {\sqrt N}}\sum_\mathbf{k}  e^{\mathrm{i}\mathbf{k}\cdot
\mathbf{R_i}}\varepsilon (\mathbf{k})\psi (\mathbf{k},t)
\end{equation}
where $\mathbf{k}$ runs over the first Brillouin zone. The
constraint (\ref{2.9}) gives
\begin{equation}\label{2.12}
m(\mathbf{k})=\varepsilon (\mathbf{k})I(\mathbf{k})
\end{equation}
after defining the normalization matrix
\begin{equation}\label{2.13}
I(\mathbf{i},\mathbf{j})=\langle [\psi (\mathbf{i},t),\psi ^\dag
(\mathbf{j},t)]_\eta \rangle ={1 \over N}\sum_\mathbf{k}
e^{\mathrm{i}\mathbf{k}\cdot
(\mathbf{R_i}-\mathbf{R_j})}I(\mathbf{k})
\end{equation}
and the m-matrix
\begin{equation}\label{2.14}
m(\mathbf{i},\mathbf{j})=\langle [J(\mathbf{i},t),\psi ^\dag
(\mathbf{j},t)]_\eta \rangle ={1 \over N}\sum_\mathbf{k}
e^{\mathrm{i}\mathbf{k}\cdot
(\mathbf{R_i}-\mathbf{R_j})}m(\mathbf{k})
\end{equation}

Since the components of $\psi (i)$ contain composite operators, the
normalization matrix $I(\mathbf{k})$ is not the identity matrix and
defines the spectral content of the excitations In fact, the use of
composite operators has the advantage of describing crossover
phenomena as the phenomena in which the weight of some operator is
shifted to another one.

It is worth noting that the normalization matrix $I(\mathbf{k})$ and
the $m(\mathbf{k})$ matrix are the lowest order generalized spectral
moments  $M^{(p)}(\mathbf{k})$ which are defined as
\begin{equation}\label{2.15}
M^{(p)}(\mathbf{k})=F.T.\left\langle {\left[ {\left(
{\mathrm{i}\partial /\partial t} \right)^p\psi (\mathbf{i},t),\psi
^\dag (\mathbf{j},t)} \right]_\eta } \right\rangle
\end{equation}
where $F.T.$ stands for the Fourier transform. The generalized
spectral moments $M^{(p)}(k)$ have the Hermiticity property
\begin{equation}\label{2.16}
M_{ab}^{(p)}(\mathbf{k})=\left( {M_{ba}^{(p)}(\mathbf{k})} \right)*
\end{equation}
since at the equilibrium $M^{(p)}(\mathbf{k})$ is time-independent.

Coming back to our original problem, the evaluation of the
expectation value $\langle A\rangle $, it is possible to use the
equation of motion $\mathrm{i}{\partial  \over {\partial t}}\psi
(i)=[\psi (i),H]$ in order to derive equations for $\langle A\rangle
$. However, the correlation functions satisfy homogeneous equations.
A convenient generalization of the concept of correlation functions
is furnished by the Green's functions (GF) which have some
advantages in the construction and solution of the equations
determining them. In particular, the two-time Green's functions
contain most of the relevant information on the properties of the
system: expectation values of the observables, excitation spectrum,
response to external perturbation, and so on. Different types of GF
can be defined; for statistical systems it is better to consider the
real time thermodynamic Green's functions where the averaging
process of the Heisenberg operators is performed over the grand
canonical ensemble.

By considering the two-time thermodynamic Green's functions
\cite{Bogoliubov_59,Zubarev_60,Zubarev_74}, let us define the causal
function
\begin{equation}\label{2.17a}
G^C(i,j)=\langle C[\psi (i)\psi ^\dag (j)]\rangle =\theta
(t_i-t_j)\langle \psi (i)\psi ^\dag (j)\rangle -\eta \theta
(t_j-t_i)\langle \psi ^\dag (j)\psi (i)\rangle
\end{equation}
the retarded and advanced functions
\begin{equation}\label{2.17b}
G^{R,A}(i,j)=\langle R,A[\psi (i)\psi ^\dag (j)]\rangle =\pm \theta
[\pm (t_i-t_j)]\langle [\psi (i),\psi ^\dag (j)]_\eta
\rangle
\end{equation}

By means of the Heisenberg equation (\ref{2.7}) and using the
decomposition (\ref{2.8}), the Green's function $G^Q(i,j)=\langle Q[
\psi (i) \psi ^\dag (j)]\rangle$, where $Q=C,\,R,\,A$, satisfies the
equation
\begin{equation}\label{2.18}
\Lambda (\partial _i)G^Q(i,j)\Lambda ^\dag
(\mathord{\buildrel{\lower3pt\hbox{$\scriptscriptstyle\leftarrow$}}\over
\partial } _j)=\Lambda (\partial _i)G_0^Q(i,j)\Lambda ^\dag
(\mathord{\buildrel{\lower3pt\hbox{$\scriptscriptstyle\leftarrow$}}\over
\partial } _j)+\langle Q[\delta J(i)\delta J^\dag (j)]\rangle
\end{equation}
where the derivative operator $\Lambda (\partial _i)$ is defined as
\begin{equation}\label{2.19}
\Lambda (\partial _i)=\mathrm{i}{\partial  \over {\partial
t_i}}-\varepsilon (-\mathrm{i}\nabla _i)
\end{equation}
and the propagator $G_0^Q(i,j)$ is defined by the equation
\begin{equation}\label{2.20}
\Lambda (\partial _i)G_0^Q(i,j)=i\delta
(t_i-t_j)I(\mathbf{i},\mathbf{j})
\end{equation}

By introducing the Fourier transform
\begin{equation}\label{2.21}
G^Q(i,j)={1 \over N}\sum_\mathbf{k}{\mathrm{i} \over {(2\pi
)}}\int_{-\infty }^{+\infty }  d\omega \kern 1pt
e^{\mathrm{i}\mathbf{k}\cdot
(\mathbf{R_i}-\mathbf{R_j})-\mathrm{i}\omega
(t_i-t_j)}G^Q(\mathbf{k},\omega )
\end{equation}
equation (\ref{2.18}) in momentum space is written as
\begin{equation}\label{2.22}
G^Q(\mathbf{k},\omega )=G_0^Q(\mathbf{k},\omega
)+G_0^Q(\mathbf{k},\omega )\Sigma^{Q*}(\mathbf{k},\omega
)G_0^Q(\mathbf{k},\omega )
\end{equation}
where the self-energy $\Sigma ^{Q*}(\mathbf{k},\omega )$ has the
expression
\begin{equation}\label{2.23}
\Sigma ^{Q*}(\mathbf{k},\omega
)=I^{-1}(\mathbf{k})B^Q(\mathbf{k},\omega )I^{-1}(\mathbf{k})
\end{equation}
with
\begin{equation}\label{2.24}
B^Q(\mathbf{k},\omega )=F.T.\langle Q[\delta J(i)\delta J^\dag
(j)]\rangle
\end{equation}

Next, we introduce the irreducible self-energy $\Sigma
^Q(\mathbf{k},\omega )$ by means of the definition
\begin{equation}\label{2.25}
\Sigma ^Q(\mathbf{k},\omega )G^Q(\mathbf{k},\omega
)=I(\mathbf{k})\Sigma ^{Q*}(\mathbf{k},\omega
)G_0^Q(\mathbf{k},\omega )
\end{equation}
Equation (\ref{2.22}) takes the form
\begin{equation}\label{2.26}
G^Q(\mathbf{k},\omega )=G_0^Q(\mathbf{k},\omega
)+G_0^Q(\mathbf{k},\omega )I^{-1}(\mathbf{k})\Sigma
^Q(\mathbf{k},\omega )G^Q(\mathbf{k},\omega )
\end{equation}
and can be formally solved as
\begin{equation}\label{2.27}
G^Q(\mathbf{k},\omega )={1 \over {\omega -\varepsilon
(\mathbf{k})-\Sigma ^Q(\mathbf{k},\omega )}}I(\mathbf{k})
\end{equation}

The formal definition (\ref{2.23}) of self-energy must be
manipulated to avoid any flowing on tautology. By noting that
$G^{Q-1}(\mathbf{k},\omega )=I^{-1}(\mathbf{k})[\omega -\varepsilon
(\mathbf{k})-\Sigma ^Q(\mathbf{k},\omega )]$ we can express $\Sigma
^Q(\mathbf{k},\omega )$ as
\begin{equation}\label{2.28}
\Sigma ^Q(\mathbf{k},\omega )=B_{irr}^Q(\mathbf{k},\omega
)I^{-1}(\mathbf{k})
\end{equation}
where $B_{irr}^Q(\mathbf{k},\omega )$ indicates the irreducible part
of the propagator $B^Q(\mathbf{k},\omega )$, given by
\begin{equation}\label{2.29}
B_{irr}^Q(\mathbf{k},\omega )={1 \over {B^{Q-1}(\mathbf{k},\omega
)+I^{-1}(\mathbf{k})G_0^Q(\mathbf{k},\omega )I^{-1}(\mathbf{k})}}
\end{equation}

We have constructed a generalized perturbative approach designed for
formulations using composite fields. Equation (\ref{2.26}) is a
Dyson-like equation and may represents the starting point for a
perturbative calculation in terms of the propagator
$G_0^Q(\mathbf{k},\omega )$. Contrarily to the usual perturbation
schemes, the calculation of the "free propagator"
$G_0^Q(\mathbf{k},\omega )$ is not an easy task and large part of
this article will be dedicated to this problem. Then, the attention
will be given to the calculation of the self-energy $\Sigma
^Q(\mathbf{k},\omega )$, and some approximate methods will be
presented. It should be noted that the computation of the two
quantities $G_0^Q(\mathbf{k},\omega )$ and $\Sigma
^Q(\mathbf{k},\omega )$ are intimately related. The total weight of
the self-energy corrections is bounded by the weight of the residual
source operator $\delta J(i)$. According to this, it can be made
smaller and smaller by increasing the components of the basis $\psi
(i)$ [e.g. by including higher-order composite operators appearing
in $\delta J(i)$]. The result of such a procedure will be the
inclusion in the energy matrix of part of the self-energy as an
expansion in terms of coupling constants multiplied by the weights
of the newly includes basis operators. In general, the enlargement
of the basis leads to a new self-energy with a smaller total weight.
However, it is necessary pointing out that this process can be quite
cumbersome and the inclusion of fully momentum and frequency
dependent self-energy corrections can be necessary to effectively
take into account low-energy and virtual processes. According to
this, one can chose a reasonable number of components for the basic
set and then use another approximation method to evaluate the
residual dynamical corrections.

\section{GF properties, spectral representation, zero-frequency
functions}

\subsection{Equations of motion}

In the previous Section we have constructed a generalized
perturbative approach based on a Dyson equation designed for
formulations using composite fields. Two quantities appear in the
Dyson equation (\ref{2.26}): the "free propagator"
$G_0^Q(\mathbf{k},\omega )$ and the self-energy $\Sigma
^Q(\mathbf{k},\omega )$. By postponing to next Sections the problem
of computing the self-energy, in this Section we concentrate on the
calculation of the Green's functions $G_0^Q(\mathbf{k},\omega )$
which constitute the building blocks of the perturbation scheme we
are trying to formulate. For the sake of simplicity we will drop the
sub index $0$ in the definition of $G_0^Q(k,\omega )$.

One fundamental aspect in a Green's function formulation is the
choice of the representation. The knowledge of the Hamiltonian and
of the operatorial algebra is not sufficient to completely specify
the GF. The GF refer to a specific representation (i.e., to a
specific choice of the Hilbert space) and this information must be
supplied as a boundary condition to the equations of motion that
alone are not sufficient to completely determine the GF. As well
known, the same system can exist in different phases according to
the external conditions; the existence of infinite inequivalent
representations \cite{Umezawa_65,Umezawa_66,Leplae_74} where the
equations of motions can be realized, allows us to pick up, among
the many possible choices, the right Hilbert space appropriate to
the physical situation under study. The use of composite operators
leads to an enlargement of the Hilbert space by the inclusion of
some unphysical states. As a consequence of this, it is difficult to
satisfy a priori all the sum rules and, in general, the symmetry
properties enjoined by the system under study. In addition, since
the representation where the operators are realized has to be
dynamically determined, the method clearly requires a process of
self-consistency.

From this discussion it is clear that fixing the representation is
not an easy task and requires special attention. In the literature
the properties of the GF are usually determined by starting from the
knowledge of the representation. Owing to the difficulties above
discussed we cannot proceed in this way. Therefore, we will derive
the general properties of the GF on the basis of the two elements we
have: the dynamics, fixed by the choice of the Hamiltonian
(\ref{2.1}), and the algebra, fixed by the choice of the basic set
(\ref{2.5}). The problem of fixing the representation will be
considered in the next Sections.

Let $\psi (i)$ be a n-component field satisfying linear equations of
motion
\begin{equation}\label{3.1}
\mathrm{i}{\partial  \over {\partial t}}\psi
_m(\mathbf{i},t)=\sum_\mathbf{j} \sum_{l=1}^n \varepsilon
_{ml}(\mathbf{i},\mathbf{j})\psi _l(\mathbf{j},t)
\end{equation}
with the energy matrix $\varepsilon (\mathbf{i},\mathbf{j})$ defined
by (\ref{2.12})-(\ref{2.14}). If the fields $\psi (i)$ are
eigenoperators of the total Hamiltonian, the equations of motion
(\ref{3.1}) are exact. There are many non-trivial realistic systems
for which it is possible to obtain a complete set of eigenoperators
of the Hamiltonian (for instance see
\cite{Mancini_05,Mancini_05b,Mancini_05a,Avella_05}). If the fields
$\psi (i)$ are not eigenoperators of $H$, the equations are
approximated; they correspond to neglecting the residual source
operator $\delta J(i)$ in the full equation of motion [see
(\ref{2.7}) and (\ref{2.8})] and all the formalism is developed with
the intention of using the propagators of these fields as a basis to
set up a perturbative scheme of calculations on the ground of the
Dyson equation (\ref{2.26}) derived in the previous Section.

By means of the field equation (\ref{3.1}) the Fourier transforms of
the various Green's functions defined by (\ref{2.16}), (\ref{2.17a})
and (\ref{2.17b}) satisfy the following equation
\begin{equation}\label{3.2}
[\omega -\varepsilon (\mathbf{k})]G^{Q(\eta )}(\mathbf{k},\omega
)=I^{(\eta )}(\mathbf{k})
\end{equation}
where the dependence on the parameter $\eta $ has been explicitly
introduced. As mentioned in Section 2, the set $\psi(i)$ can be
fermionic or bosonic and the parameter $\eta$ generally takes the
value $\eta =1$ ($\eta =-1$) for a  fermionic (bosonic) set $\psi
(i)$. The three Green's functions $G^C$, $G^R$ and $G^A$ satisfy the
same equation of motion which alone is not sufficient and must be
supplemented by other equations. Indeed, the GF are determined by
solving a first order differential equation of motion, thereby the
GF are given only within an arbitrary constant of integration. The
retarded and advanced GF can be completely determined because the
factor $\theta [\pm (t_i-t_j)]$ provides the boundary condition:
$G^{R,A}(i,j)=0\quad for\;t_i=t_j\mp \delta $. The determination of
the causal GF is not so immediate. The most general solution of
equation (3.2) is
\begin{equation}\label{3.3}
G^{Q(\eta )}(\mathbf{k},\omega )=\sum_{l=1}^n  \left\{ {P\left(
{{{\sigma ^{(l,\eta )}(k)} \over {\omega -\omega _l(\mathbf{k})}}}
\right)-\mathrm{i}\pi \delta [\omega -\omega
_l(\mathbf{k})]g^{(l,\eta )Q}(\mathbf{k})} \right\}
\end{equation}
where $\omega _l(\mathbf{k})$ are the eigenvalues of the
$\varepsilon (\mathbf{k})$, $\sigma ^{(l)}(\mathbf{k})$ are defined
by
\begin{equation}\label{3.4}
\sigma _{ab}^{(l)}(\mathbf{k})=\sum_{c=1}^n  \Omega
_{al}(\mathbf{k})\Omega
_{lc}^{-1}(\mathbf{k})I_{cb}(\mathbf{k})\quad \quad \quad
a,b=1,\ldots,n
\end{equation}
$\Omega (\mathbf{k})$ is the $n\times n$ matrix, whose columns are
the eigenvectors of $\varepsilon (\mathbf{k})$. We note that while
the spectral density matrix $\sigma ^{(l)}(\mathbf{k})$ is
completely determined by the energy $\varepsilon (\mathbf{k})$ and
normalization $I^{(\eta )}(\mathbf{k})$ matrices, the matrix
$g^{(l,\eta )Q}(\mathbf{k})$ is not fixed by the equations of motion
and must be determined by means of the boundary conditions. $P$
represents the principal value.

By recalling the retarded and advanced nature of $G^{R,A(\eta
)}(i,j)$ it is immediate to see that
\begin{equation}\label{3.5}
g^{(l,\eta )R}(\mathbf{k})=-g^{(l,\eta )A}(\mathbf{k})=\sigma
^{(l,\eta )}(\mathbf{k})
\end{equation}
Then, the retarded and advanced functions are completely determined
in terms of the matrices $\varepsilon (\mathbf{k})$ and $I^{(\eta
)}(\mathbf{k})$. As well known, as functions of $\omega$ the
$G^{R,A(\eta )}(\mathbf{k},\omega )$ are analytic in the upper and
lower half-planes, respectively.

The determination of $g^{(l,\eta )C}(k)$ requires more work. From
the definitions (\ref{2.16}), (\ref{2.17a}) and  (\ref{2.17b}) we
can derive the following exact relations
\begin{equation}\label{3.6a}
G^{R(\eta )}(i,j)+G^{A(\eta )}(i,j)=2G^{C(\eta )}(i,j)-\langle [\psi
(i),\psi ^\dag (j)]_{-\eta }\rangle
\end{equation}
\begin{equation}\label{3.6b}
G^{R(\eta )}(i,j)-G^{A(\eta )}(i,j)=\langle [\psi (i),\psi ^\dag
(j)]_\eta \rangle
\end{equation}
where there appear the two correlation functions
\begin{equation}\label{3.7}
C_{\psi \psi ^\dag }(i,j)=\langle \psi (i)\psi ^\dag (j)\rangle
\quad \quad \quad C_{\psi ^\dag \psi }(i,j)=\langle \psi ^\dag
(j)\psi (i)\rangle
\end{equation}
The definition of $C_{\psi ^\dag \psi }(i,j)$ must be interpreted at
level of matrix elements: \\ \noindent $C_{\psi ^\dag \psi
;ba}(i,j)=\langle \psi _b^\dag (j)\psi _a(i)\rangle $; no scalar
product is intended, neither in the spin space. These functions are
defined for all real times and the equations of motion they obey
contain no inhomogeneous terms involving a delta function. Indeed,
by means of the equations of motion (\ref{3.1}) the Fourier
transform of these correlation functions satisfy the homogeneous
equations
\begin{equation}\label{3.8}
[\omega -\varepsilon (\mathbf{k})]C_{\psi \psi ^\dag
}(\mathbf{k},\omega )=0\quad \quad \quad [\omega -\varepsilon
(\mathbf{k})]C_{\psi ^\dag \psi }(\mathbf{k},\omega )=0
\end{equation}
These equations tell us that the Fourier transforms of the
correlation functions are zero unless the frequency $\omega $ is
equal to one of  the energy levels $\omega _l(\mathbf{k})$ of the
system. The solutions of (\ref{3.8}) have the general form
\begin{equation}\label{3.9a}
C_{\psi \psi ^\dag }(\mathbf{k},\omega )=\sum_{l=1}^n  \delta
[\omega -\omega _l(\mathbf{k})]c_{_{\psi \psi ^\dag
}}^{(l)}(\mathbf{k})
\end{equation}
\begin{equation}\label{3.9b}
C_{\psi ^\dag \psi }(\mathbf{k},\omega )=\sum_{l=1}^n  \delta
[\omega -\omega _l(\mathbf{k})]c_{_{\psi ^\dag \psi
}}^{(l)}(\mathbf{k})
\end{equation}
with the momentum-dependent Fourier components $c_{_{\psi \psi ^\dag
}}^{(l)}(\mathbf{k})$ and $c_{_{\psi ^\dag \psi
}}^{(l)}(\mathbf{k})$ to be determined.

We now recall the Kubo-Martin-Schwinger (KMS) relation
\begin{equation}\label{3.10}
\langle A(t)B(t)\rangle =\langle B(t)A(t+i\beta )\rangle
\end{equation}
where $A(t)$ and $B(t)$ are Heisenberg operators at time $t$. This
relation implies that the Fourier transforms $C_{\psi \psi ^\dag
}(\mathbf{k},\omega )$ and $C_{\psi ^\dag \psi }(\mathbf{k},\omega
)$ are related
\begin{equation}\label{3.11}
C_{\psi ^\dag \psi }(\mathbf{k},\omega )=e^{-\beta \omega }C_{\psi
\psi ^\dag }(\mathbf{k},\omega )
\end{equation}
and the $\eta-$commutator $\langle [\psi (i),\psi ^\dag (j)]_\eta
\rangle $ can be expressed in terms of the correlation function as
\begin{equation}\label{3.12}
\langle [\psi (i),\psi ^\dag (j)]_\eta \rangle ={1 \over
N}\sum_\mathbf{k} {1 \over {2\pi }}\int  \kern 1pt d\omega
\,e^{\mathrm{i}\mathbf{k}\cdot
(\mathbf{i}-\mathbf{j})-\mathrm{i}\omega (t_i-t_j)}[1+\eta e^{-\beta
\omega }]C_{\psi \psi ^\dag }(\mathbf{k},\omega )
\end{equation}

By putting (\ref{3.12}) into (\ref{3.6a}) and (\ref{3.6b})  and by
taking into account Eqs.~(\ref{3.3}), (\ref{3.5}) and (\ref{3.9a})
we obtain
\begin{equation}\label{3.13a}
\sum_{l=1}^n  \delta [\omega -\omega _l(\mathbf{k})]\left\{
{g^{(l,\eta )C}(\mathbf{k})-{1 \over {2\pi }}[1-\eta e^{-\beta
\omega }]c_{_{\psi \psi ^\dag }}^{(l)}(\mathbf{k})} \right\}=0
\end{equation}
\begin{equation}\label{3.13b}
\sum_{l=1}^n  \delta [\omega -\omega _l(\mathbf{k})]\left\{ {\sigma
^{(l,\eta )}(\mathbf{k})-{1 \over {2\pi }}[1+\eta e^{-\beta \omega
}]c_{_{\psi \psi ^\dag }}^{(l)}(\mathbf{k})} \right\}=0
\end{equation}

The solution of Eqs.~(\ref{3.13a}) and (\ref{3.13b}) is remarkably
different according to the value of the parameter $\eta $ and we
shall treat separately the two cases.

\subsection{Fermionic fields}

For the case of fermionic fields it is convenient to choose $\eta
=1$. Then, the solution of Eqs.~(\ref{3.13a}) and (\ref{3.13b}) is
\begin{align}\label{3.14}
&c^{(l)}(\mathbf{k})=\pi \left[ {1+\tanh \left( {{{\beta \omega
_l(\mathbf{k})} \over 2}} \right)} \right]\sigma
^{(l,+1)}(\mathbf{k})\\
&g^{(l,+1)C}(\mathbf{k})=\tanh \left( {{{\beta \omega
_l(\mathbf{k})} \over 2}} \right)\sigma ^{(l,+1)}(\mathbf{k})
\end{align}

By putting (\ref{3.5}) and (\ref{3.14}) into (\ref{3.3}),
(\ref{3.9a}) and (\ref{3.9b}) we get the following general
expressions for the Green's functions and correlation functions
\begin{equation}\label{3.15a}
G^{R,A(+1)}(\mathbf{k},\omega )=\sum_{l=1}^n  {{\sigma
^{(l,+1)}(\mathbf{k})} \over {\omega -\omega _l(\mathbf{k})\pm
\mathrm{i}\delta }}
\end{equation}
\begin{equation}\label{3.15b}
G^{C(+1)}(\mathbf{k},\omega )=\sum_{l=1}^n  \sigma
^{(l,+1)}(\mathbf{k})\left[ {{{1-f_\mathrm{F}[\omega
_l(\mathbf{k})]} \over {\omega -\omega
_l(\mathbf{k})+\mathrm{i}\delta }}+{{f_\mathrm{F}[\omega
_l(\mathbf{k})]} \over {\omega -\omega
_l(\mathbf{k})-\mathrm{i}\delta }}} \right]
\end{equation}
\begin{equation}\label{3.15c}
C_{\psi \psi ^\dag }(\mathbf{k},\omega )=\pi \sum_{l=1}^n  \delta
[\omega -\omega _l(\mathbf{k})]\left[ {1+\tanh \left( {{{\beta
\omega _l(\mathbf{k})} \over 2}} \right)} \right]\sigma
^{(l,+1)}(\mathbf{k})
\end{equation}
\begin{equation}\label{3.15d}
C_{\psi ^\dag \psi }(\mathbf{k},\omega )=\pi \sum_{l=1}^n  \delta
[\omega -\omega _l(\mathbf{k})]\left[ {1-\tanh \left( {{{\beta
\omega _l(\mathbf{k})} \over 2}} \right)} \right]\sigma
^{(l,+1)}(\mathbf{k})
\end{equation}
where $f_\mathrm{F}(\omega )$ is the Fermi distribution function:
$f_\mathrm{F}(\omega )={1 \over {e^{\beta \omega }+1}}$.

By recalling that all the fermionic energies are shifted by the
chemical potential, the locus in the $k$-space, defined by $\omega
_l(k)=0$, will define the Fermi surface. By looking at (3.15b) we
can see that the imaginary part of the causal GF vanishes on the
Fermi surface. In the fermionic case the right procedure of
calculation is to start from the retarded (advanced) GF, and then to
compute the other Green's functions and correlation functions by
means of the relations
\begin{equation}\label{3.17a}
\Re [G^{C(+1)}(\mathbf{k},\omega )]=\Re
[G^{R,A(+1)}(\mathbf{k},\omega )]
\end{equation}
\begin{equation}\label{3.17b}
\Im [G^{C(+1)}(\mathbf{k},\omega )]=\pm \tanh \left( {{{\beta \omega
} \over 2}} \right)\Im [G^{R,A(+1)}(\mathbf{k},\omega )]
\end{equation}
\begin{equation}\label{3.17c}
C_{\psi \psi ^\dag }(\mathbf{k},\omega )=\mp \left[ {1+\tanh \left(
{{{\beta \omega } \over 2}} \right)} \right]\Im
[G^{R,A(+1)}(\mathbf{k},\omega )]
\end{equation}

We note the dispersion relations
\begin{align}\label{3.18}
&\Re [G^{R,A(+1)}(\mathbf{k},\omega )]=\mp {1 \over \pi
}P\int_{-\infty }^{+\infty }  d\omega \;{1 \over {\omega -\omega
}}\Im [G^{R,A(+1)}(\mathbf{k},\omega )]\\
&\Re [G^{C(+1)}(\mathbf{k},\omega )]=-{1 \over \pi }P\int_{-\infty
}^{+\infty }  d\omega \;{1 \over {\omega -\omega }}\coth \left(
{{{\beta \omega } \over 2}} \right)\Im [G^{C(+1)}(\mathbf{k},\omega
)]
\end{align}
By introducing the spectral function
\begin{equation}\label{3.19}
\rho ^{(+1)}(\mathbf{k},\omega )=\sum_{l=1}^n  \delta [\omega
-\omega _l(\mathbf{k})]\sigma ^{(l,+1)}(\mathbf{k})=\mp {1 \over \pi
}\Im [G^{R,A(+1)}(\mathbf{k},\omega )]
\end{equation}
we can establish the spectral representation
\begin{equation}\label{3.20a}
G^{R,A(+1)}(\mathbf{k},\omega )=\int_{-\infty }^{+\infty }  d\omega'
\;{{\rho ^{(+1)}(\mathbf{k},\omega' )} \over {\omega -\omega' \pm
\mathrm{i}\delta}}
\end{equation}
\begin{equation}\label{3.21b}
G^{C(+1)}(\mathbf{k},\omega )=\int_{-\infty }^{+\infty }  d\omega'
\;\rho ^{(+1)}(\mathbf{k},\omega' )\left[ {{{1-f_\mathrm{F}(\omega'
)} \over {\omega -\omega' +\mathrm{i}\delta
}}+{{f_\mathrm{F}(\omega' )} \over {\omega -\omega'
-\mathrm{i}\delta }}} \right]
\end{equation}

\subsection{Bosonic fields}

For the case of bosonic fields it is convenient to choose $\eta
=-1$. For any given momentum $\mathbf{k}$ we can always write
\begin{equation}\label{3.22}
\omega _l(\mathbf{k})=\left\{
\begin{array}{l}
=0 \quad \text{for} \;l \in A(\mathbf{k})\subseteq N=\{ 1,\ldots,n\}\\
\ne 0 \quad \text{for} \; l \in B(\mathbf{k})=N-A(\mathbf{k})
\end{array}
\right.
\end{equation}

Obviously, $A(\mathbf{k})$ can also be the empty set (i.e.,
$A(\mathbf{k})=\emptyset $ and $B(\mathbf{k})=N$). For $l\in
B(\mathbf{k})$ the solution of (\ref{3.13a}) and (\ref{3.13b}) is
\begin{align}\label{3.23}
&c_{\psi \psi ^\dag}^{(l)}(\mathbf{k})=\pi \left[ {1+\coth \left(
{{{\beta \omega _l(\mathbf{k})} \over 2}} \right)} \right]\sigma
^{(l,-1)}(\mathbf{k})\quad \quad \forall l\in
B(\mathbf{k})\\
&g^{(l,-1)C}(\mathbf{k})=\coth \left( {{{\beta \omega
_l(\mathbf{k})} \over 2}} \right)\sigma ^{(l,-1)}(\mathbf{k})\quad
\quad \quad \quad \forall l\in B(\mathbf{k})
\end{align}
For $l\in A(\mathbf{k})$ the Fourier coefficients $c_{_{\psi \psi
^\dag }}^{(l)}(\mathbf{k})$ and  $g^{(l,-1)C}(\mathbf{k})$ cannot be
determined from Eqs.~(\ref{3.13b}). It is convenient to introduce
the function $\Gamma (\mathbf{k})$
\begin{equation}\label{3.24}
\Gamma (\mathbf{k})={1 \over {2\pi }}\sum_{l\in A(\mathbf{k})}
c_{_{\psi \psi ^\dag }}^{(l)}(\mathbf{k})={1 \over 2}\sum_{l\in
A(\mathbf{k})} g^{(l,-1)C}(\mathbf{k})
\end{equation}
By considering that from (\ref{3.13b}) and (\ref{3.9a})
\begin{equation}\label{3.25}
\mathop {\lim }_{\omega \to 0}[1-e^{-\beta \omega }]C_{\psi \psi
^\dag }(\mathbf{k},\omega )=2\pi \delta (\omega )\sum_{l\in
A(\mathbf{k})} \sigma ^{(l,-1)}(\mathbf{k})
\end{equation}
we must distinguish two situations:
\begin{enumerate}
 \item If  $\sum_{l\in A(\mathbf{k})} \sigma ^{(l,-1)}(\mathbf{k})=0$, then
\begin{equation}\label{3.26a}
\sum_{l=1}^n  \delta [\omega -\omega _l(\mathbf{k})]c_{_{\psi \psi
^\dag }}^{(l)}(\mathbf{k})=2\pi \delta (\omega )\Gamma
(\mathbf{k})+2\pi \sum_{l\in B(\mathbf{k})} \delta [\omega -\omega
_l(\mathbf{k})]{{e^{\beta \omega _l(\mathbf{k})}} \over {e^{\beta
\omega _l(\mathbf{k})}-1}}\sigma ^{(l,-1)}(\mathbf{k})
\end{equation}
\begin{equation}\label{3.26b}
\sum_{l=1}^n  \delta [\omega -\omega _l(\mathbf{k})]c_{_{\psi ^\dag
\psi }}^{(l)}(\mathbf{k})=2\pi \delta (\omega )\Gamma
(\mathbf{k})+2\pi \sum_{l\in B(\mathbf{k})} \delta [\omega -\omega
_l(\mathbf{k})]{1 \over {e^{\beta \omega _l(\mathbf{k})}-1}}\sigma
^{(l,-1)}(\mathbf{k})
\end{equation}
and
\begin{equation}\label{3.27}
C_{\psi \psi ^\dag }(\mathbf{k},0)=C_{\psi ^\dag \psi
}(\mathbf{k},0)
\end{equation}
 \item If $\sum_{l\in A(\mathbf{k})} \sigma ^{(l,-1)}(\mathbf{k})\ne 0$ but finite,
then in order to satisfy (\ref{3.25}) $C_{\psi \psi ^\dag
}(\mathbf{k},\omega )$ must have a  singularity of the type ${1
\over \omega }$ in the limit $\omega \to 0$. In fact
\begin{equation}\label{3.28}
(1-e^{-\beta \omega }){1 \over {\beta \omega }}=(\beta \omega -{1
\over 2}\beta ^2\omega ^2+\ldots){1 \over {\beta \omega }}=1-{1
\over 2}\beta \omega +\ldots
\end{equation}
Then
\begin{equation}\label{3.29a}
\sum_{l=1}^n  \delta [\omega -\omega _l(\mathbf{k})]c_{_{\psi \psi
^\dag }}^{(l)}(\mathbf{k})=2\pi \delta (\omega )\Gamma
(\mathbf{k})+2\pi \sum_{l=1}^n \delta [\omega -\omega
_l(\mathbf{k})]{{e^{\beta \omega _l(\mathbf{k})}} \over {e^{\beta
\omega _l(\mathbf{k})}-1}}\sigma ^{(l,-1)}(\mathbf{k})
\end{equation}
\begin{equation}\label{3.29b}
\sum_{l=1}^n  \delta [\omega -\omega _l(\mathbf{k})]c_{_{\psi ^\dag
\psi }}^{(l)}(\mathbf{k})=2\pi \delta (\omega )\Gamma
(\mathbf{k})+2\pi \sum_{l=1}^n \delta [\omega -\omega
_l(\mathbf{k})]{1 \over {e^{\beta \omega _l(\mathbf{k})}-1}}\sigma
^{(l,-1)}(\mathbf{k})
\end{equation}
and
\begin{equation}\label{3.30}
C_{\psi \psi ^\dag }(\mathbf{k},0)-C_{\psi ^\dag \psi
}(\mathbf{k},0)=\sum_{l\in A(\mathbf{k})} \sigma
^{(l,-1)}(\mathbf{k})
\end{equation}
\end{enumerate}

It is clear from (\ref{3.29a}) and (\ref{3.29b}) that the situation
where $\sum_{l\in A(\mathbf{k})} \sigma ^{(l,-1)}(\mathbf{k})\ne 0$
leads to a situation in which for $l\in A(\mathbf{k})$ the Fourier
coefficients $c_{_{\psi \psi ^\dag }}^{(l)}(\mathbf{k})$ and
$c_{_{\psi ^\dag \psi }}^{(l)}(\mathbf{k})$ diverge as $[\beta
\omega _l(\mathbf{k})]^{-1}$. Since the correlation function in
direct space must be finite, at finite temperature this is
admissible only in the thermodynamic limit and if the dispersion
relation $\omega _l(\mathbf{k})$ is such that the divergence in
momentum space is integrable and the corresponding correlation
function in real space remains finite. For finite systems and for
infinite systems where the divergence is not integrable we must have
$\sum_{l\in A(\mathbf{k})}\sigma ^{(l,-1)}(\mathbf{k})=0$. The
calculation of the spectral density matrices $\sigma
^{(l,-1)}(\mathbf{k})$ it not a simple dynamical problem, but
requires the self-consistent calculation of some expectation values,
where the boundary condition and the choice of the representation
play a crucial role. A finite value of $\sum_{l\in A(\mathbf{k})}
\sigma ^{(l,-1)}(\mathbf{k})$ is generally related to the presence
of long-range order and the previous statement is nothing but the
Mermin-Wagner theorem \cite{Mermin_66}.

Summarizing, by using (\ref{3.5}) and by putting (\ref{3.26a}) and
(\ref{3.26b}) into (\ref{3.13a}), for finite systems and $T\ne 0$,
we get the following general expressions for the GF and correlations
function
\begin{equation}\label{3.31a}
G^{R,A(-1)}(\mathbf{k},\omega )=\sum_{l=1}^n {{\sigma
^{(l,-1)}(\mathbf{k})} \over {\omega -\omega _l(\mathbf{k})\pm
\mathrm{i}\delta }}
\end{equation}
\begin{align}\label{3.31b}
&G^{C(-1)}(\mathbf{k},\omega )=\Gamma (\mathbf{k})\left[ {{1 \over
{\omega +\mathrm{i}\delta }}-{1 \over {\omega -\mathrm{i}\delta }}}
\right]\nonumber\\
&+\sum_{l\in B(\mathbf{k})} \sigma ^{(l,-1)}(\mathbf{k})\left[
{{{1+f_\mathrm{B}(\omega )} \over {\omega -\omega
_l(\mathbf{k})+\mathrm{i}\delta }}-{{f_\mathrm{B}(\omega )} \over
{\omega -\omega _l(\mathbf{k})-\mathrm{i}\delta }}} \right]
\end{align}
\begin{equation}\label{3.31c}
C_{\psi \psi ^\dag }(\mathbf{k},\omega )=2\pi \Gamma
(\mathbf{k})\delta (\omega )+\pi \sum_{l\in B(\mathbf{k})}  \delta
[\omega -\omega _l(\mathbf{k})]\left[ {1+\coth \left( {{{\beta
\omega _l(\mathbf{k})} \over 2}} \right)} \right]\sigma
^{(l,-1)}(\mathbf{k})
\end{equation}
\begin{equation}\label{3.31d}
C_{\psi ^\dag \psi }(\mathbf{k},\omega )=2\pi \Gamma
(\mathbf{k})\delta (\omega )-\pi \sum_{l\in B(\mathbf{k})}  \delta
[\omega -\omega _l(\mathbf{k})]\left[ {1-\coth \left( {{{\beta
\omega _l(\mathbf{k})} \over 2}} \right)} \right]\sigma
^{(l,-1)}(\mathbf{k})
\end{equation}
with the condition that $\sum_{l\in A(\mathbf{k})}  \sigma
^{(l,-1)}(\mathbf{k})=0$. $f_\mathrm{B}(\omega )$ is the Bose
distribution function: $f_\mathrm{B}(\omega )={1 \over {e^{\beta
\omega }-1}}$. It is possible to have $\sigma
^{(l,-1)}(\mathbf{k})\ne 0\quad for\quad l\in A(\mathbf{k})$ only
for infinite systems and if the divergence is integrable.

The previous formulas show that when zero-energy modes are present a
zero-frequency singularity appears in the correlation function
$C_{\psi \psi ^\dag }(\mathbf{k},\omega )$ and in the imaginary part
of the causal function $G^{C(-1)}(\mathbf{k},\omega )$. Such
singularity does not contribute to the retarded and advanced GF.
Then, in the bosonic case the right procedure of calculation is to
start from the causal GF, and compute the other GF by means of the
relations
\begin{equation}\label{3.34a}
\Re [G^{R,A(-1)}(\mathbf{k},\omega )]=\Re
[G^{C(-1)}(\mathbf{k},\omega )]
\end{equation}
\begin{equation}\label{3.34b}
\Im [G^{R,A(-1)}(\mathbf{k},\omega )]=\pm \tanh \left( {{{\beta
\omega } \over 2}} \right)\Im [G^{C(-1)}(\mathbf{k},\omega )]
\end{equation}
\begin{equation}\label{3.34c}
C_{\psi \psi ^\dag }(\mathbf{k},\omega )=-\left[ {1+\tanh \left(
{{{\beta \omega } \over 2}} \right)} \right]\Im
[G^{C(-1)}(\mathbf{k},\omega )]
\end{equation}

We note the dispersion relations
\begin{align}\label{3.35}
&\Re [G^{R,A(-1)}(\mathbf{k},\omega )]=\mp {1 \over \pi
}P\int_{-\infty }^{+\infty }  d\omega \;{1 \over {\omega -\omega
}}\Im [G^{R,A(-1)}(\mathbf{k},\omega )]\\
&\Re [G^{C(-1)}(\mathbf{k},\omega )]=-{1 \over \pi }P\int_{-\infty
}^{+\infty }  d\omega \;{1 \over {\omega -\omega }}\tanh \left(
{{{\beta \omega } \over 2}} \right)\Im [G^{C(-1)}(\mathbf{k},\omega
)]
\end{align}

Also in the bosonic case we can introduce a spectral function
\begin{equation}\label{3.36}
\rho ^{(-1)}(\mathbf{k},\omega )=\sum_{l=1}^n  \delta [\omega
-\omega _l(\mathbf{k})]\sigma ^{(l,-1)}(\mathbf{k})=\mp {1 \over \pi
}\Im [G^{R,A(-1)}(\mathbf{k},\omega )]
\end{equation}

However, the zero-frequency function $\Gamma (\mathbf{k})$ does not
contribute to $\rho ^{(-1)}(\mathbf{k},\omega )$ and a spectral
representation can be established only for the retarded (advanced)
GF
\begin{equation}\label{3.37}
G^{R,A(-1)}(\mathbf{k},\omega )=\int_{-\infty }^{+\infty }  d\omega'
\;{{\rho ^{(-1)}(\mathbf{k},\omega' )} \over {\omega -\omega' \pm
\mathrm{i}\delta }}
\end{equation}

For the bosonic causal GF a spectral representation exists only when
$\Gamma (\mathbf{k})=0$:
\begin{equation}\label{3.38}
G^{C(-1)}(\mathbf{k},\omega )=\int_{-\infty }^{+\infty } d\omega'
\;\rho ^{(-1)}(\mathbf{k},\omega' )\left[ {{{1+f_\mathrm{B}(\omega'
)} \over {\omega -\omega' +\mathrm{i}\delta
}}-{{f_\mathrm{B}(\omega' )} \over {\omega -\omega'
-\mathrm{i}\delta }}} \right]
\end{equation}

\subsection{Sum rules and some useful relations}

Coming back to a generic value of $\Gamma (\mathbf{k})$, we note
that from the definition (\ref{3.4}) the following sum rule can be
derived
\begin{equation}\label{3.39}
\int_{-\infty }^{+\infty }  d\omega \kern 1pt \rho ^{(\eta
)}(\mathbf{k},\omega )=\sum_{l=1}^n  \sigma ^{(l,\eta
)}(\mathbf{k})=I^{(\eta )}(\mathbf{k})
\end{equation}

This is a particular case of a general sum rule. From the results
(\ref{3.15c}), (\ref{3.15d}), (\ref{3.31c}) and (\ref{3.31d}) we
obtain
\begin{equation}\label{3.40}
\langle [\psi (i),\psi ^\dag (j)]_\eta \rangle =\sum_{l=1}^n  {1
\over N}\sum_\mathbf{k}  \,e^{\mathrm{i}\mathbf{k}\cdot
(\mathbf{i}-\mathbf{j})-\mathrm{i}\omega
_l(\mathbf{k})(t_i-t_j)}\sigma ^{(l,\eta )}(\mathbf{k})
\end{equation}
Recalling the expression (\ref{2.15}) of the generalized spectral
moment $M^{(p,\eta )}(\mathbf{k})$ we immediately have
\begin{equation}\label{3.41}
\int_{-\infty }^{+\infty }  d\omega \kern 1pt \omega ^p\rho ^{(\eta
)}(\mathbf{k},\omega )=\sum_{l=1}^n  \,\omega _l(\mathbf{k})^p\sigma
^{(l,\eta )}(\mathbf{k})=M^{(p,\eta )}(\mathbf{k})
\end{equation}

Some interesting results can be obtained by noting that the
correlation function $C_{\psi \psi ^\dag }(i,j)=\langle \psi (i)\psi
^\dag (j)\rangle $ and the energy matrix $\varepsilon
(\mathbf{i},\mathbf{j})$ do not depend on $\eta $. As mentioned
above, once we have chosen a basic set $\{ \psi (i)\} $, fermionic
or bosonic, it is a only a question of convenience to choose$\eta
=1$ or $\eta =-1$. Let us consider the case of a bosonic set $\{
\psi (i)\} $ and let us suppose to perform two series of
calculations: one with $\eta =-1$ and another with $\eta =1$. Then,
it is immediate to obtain the following relations:
\begin{enumerate}
 \item by equating (\ref{3.15c}) and (\ref{3.31c}) we obtain
\begin{equation}\label{3.42a}
\Gamma (\mathbf{k})={1 \over 2}\sum_{l\in A(\mathbf{k})}  \sigma
^{(l,+1)}(\mathbf{k})
\end{equation}
\begin{equation}\label{3.42b}
\sigma ^{(l,-1)}(\mathbf{k})=\tanh \left( {{{\beta \omega
_l(\mathbf{k})} \over 2}} \right)\sigma ^{(l,+1)}(\mathbf{k})\quad
\quad \forall \;l\in B(\mathbf{k})
\end{equation}
 \item by equating (\ref{3.17c}) and (\ref{3.34c}) we obtain
\begin{equation}\label{3.43}
\Im [G^{C(-1)}(\mathbf{k},\omega )]=\pm \Im
[G^{R,A(+1)}(\mathbf{k},\omega )]
\end{equation}
 \item from the sum rule (\ref{3.39}) and by means of (\ref{3.42b}) and
(\ref{3.4}) we obtain
\begin{equation}\label{3.44}
I_{ab}^{(-1)}(\mathbf{k})=\sum_{l=1}^n  \tanh \left( {{{\beta \omega
_l(\mathbf{k})} \over 2}} \right)\sum_{c=1}^n  \Omega
_{al}(\mathbf{k})\Omega
_{lc}^{-1}(\mathbf{k})I_{cb}^{(+1)}(\mathbf{k})
\end{equation}
\end{enumerate}

We see that the general structure of the GF is remarkably different
according to the statistics. For fermionic composite fields (i.e.,
when it is natural to choose $\eta =1$) all the Green's functions
and correlation functions are completely determined. The
zero-frequency function $\Gamma (\mathbf{k})$, defined on the Fermi
surface $\omega _l(\mathbf{k})=\mu$, contributes to the spectral
function $\rho ^{(+1)}(\mathbf{k},\omega )$ (see \ref{3.19}), it is
directly related to the spectral density functions $\sigma
^{(l,+1)}(\mathbf{k})$ by means of equation (\ref{3.42a}), and its
calculation does not require more information. Also, it does not
contribute to the imaginary part of the causal GF. For bosonic
composite fields (i.e., when it is natural to choose $\eta=-1$) the
retarded and advanced GF are completely determined, but the causal
GF and the correlation function depend on the zero-frequency
function $\Gamma (\mathbf{k})$, defined on the surface  $\omega
_l(\mathbf{k})=0$. It is now clear that the causal and retarded
(advanced) GF contain different information and that the right
procedure of calculation is controlled by the statistics. In
particular, in the case of bosonic fields (i.e., for $\eta =-1$) one
must start from the causal function and then use (3.34) to compute
the other GF. On the contrary, for fermionic fields (i.e., for $\eta
=1$) the right procedure for computing the correlation function
requires first the calculation of the retarded (advanced) function
and then the use (\ref{3.17a}), (\ref{3.17b}) and (\ref{3.17c}) to
compute the other GF. Moreover, it is worth noting that $\Gamma
(\mathbf{k})$ is undetermined within the bosonic sector (i.e., $\eta
=-1$). It is true that $\Gamma (\mathbf{k})$ could be computed by
considering an anticommutating algebra: remaining in the bosonic
sector we make the choice $\eta =1$ and $\Gamma (\mathbf{k})$ can be
calculated by (\ref{3.42a}) or equivalently by means of the
following relation $\Gamma (\mathbf{k})={1 \over 2}\mathop {\lim
}_{\omega \to 0}\omega G^{C(+1)}(\mathbf{k},\omega )$ which can be
easily obtained from (\ref{3.31b}). However, the calculation of the
$\sigma ^{(l,+1)}(\mathbf{k})$ requires the calculation of the
normalization matrix $I^{(+1)}(\mathbf{k})$ that, for bosonic
fields, generates unknown momentum dependent correlation functions
whose determination can be very cumbersome as requires, at least in
principle, the self-consistent solution of the integral equations
connecting them to the corresponding Green's functions. In practice,
also for simple, but anyway composite, bosonic fields the $\Gamma
(\mathbf{k})$ remains undetermined and other methods should be used.
Similar methods, like the use of the relaxation function
\cite{Kubo_57}, would lead to the same problem.

Actually, all issues related to $\Gamma (\mathbf{k})$ have a natural
playground in dealing with the ergodicity of the dynamics under
investigation. More detail on this topic can be found in a
manuscript in this same issue \cite{Avella_06}.

The formulation given in this Section needs some modifications in
the case of zero temperature. In particular, Eqs.~(\ref{3.13a}) and
(\ref{3.13b}) are not applicable and we must proceed in a different
way. After a straightforward derivation \cite{Mancini_00}, it is
immediate to see that the limit $T\to 0$ of the expressions
(\ref{3.15c}), (\ref{3.15d}), (\ref{3.31c}) and (\ref{3.31d}) gives
the right result.

\section{A self-consistent scheme}

As stressed in Section 1, in the study of highly interacting
systems, where traditional perturbative calculations in terms of the
noninteracting fields fail, a way to reconcile the powerful
perturbation theory with the presence of complex and/or strong
interactions is to describe the system in terms of a new set of
fields, composite operators, generated by the interactions
themselves. These field operators undoubtedly constitute a better
starting point: they appear as the final result of the modifications
imposed by the interactions on the original particles and contain,
by the very beginning, the effects of the correlations.   Once a
choice of composite fields has been made, the relative Green's
function formalism can be set up, as illustrated in Section 2, where
a generalized Dyson equation has been derived. On this basis one can
construct a non-standard perturbation formalism where the basic
ingredients are the propagators of a subset of the fundamental
basis, satisfying linear equations of motion [cfr.~(\ref{3.1})]. By
means of the equations of motion and by using the boundary
conditions related to the definitions of the various Green's
functions we have been able to derive explicit expressions for these
latter [cfr.~(\ref{3.15c}), (\ref{3.15d}), (\ref{3.31c}) and
(\ref{3.31d})]. However, these expressions can only determine the
functional dependence; the knowledge of the GF is not fully achieved
yet. The reason is that the algebra of the field is not canonical.
As a consequence, the inhomogeneous terms $I^{(\eta )}(\mathbf{k})$
in the equations of motion (\ref{3.2}) and the energy matrix
$\varepsilon (\mathbf{k})$ contain some unknown static correlation
functions, correlators, that have to be self-consistently
calculated. Three serious problems arise with the study of the
Green's functions: (a) the calculation of some parameters expressed
as correlation functions of field operators not belonging the chosen
basis; (b) the appearance of some zero-frequency constants (ZFC) and
their determination; (c) the problem of fixing the representation
where the Green's functions are formulated.

In the Composite Operator Method \cite{Mancini_00,Mancini_04} (COM)
the three problems (a), (b) and (c) are not considered separately
but they are all connected in one self-consistent scheme. The main
idea is that fixing the values of the unknown parameters and of the
ZFC implies to put some constraints on the representation where the
GF are realized. As the determination of this representation is not
arbitrary, it is clear that there is no freedom in fixing these
quantities. They must assume values compatible with the dynamics and
with the right representation. Which is the right representation?

From the algebra it is possible to derive several relations among
the operators. We will call algebra constraints (AC) all possible
relations among the operators dictated by the algebra. This set of
relations valid at microscopic level must be satisfied also at
macroscopic level, when expectations values are considered. Also, we
note that, in general, the Hamiltonian has some symmetry properties
(i.e. rotational invariance in coordinate and spin space, phase
invariance, gauge invariance,\ldots). These symmetries generate a
set of relations among the matrix elements: the Ward-Takahashi
identities \cite{Ward_50,Takahashi_57} (WT).

Now, certainly the right representation must be the one where the
relations among the operators and the conservation laws are
maintained when expectation values are taken; in other words, all
the AC and WT are preserved. By imposing these conditions we obtain
a set of self-consistent equations that will fix the unknown
correlators, the ZFC and the right representation at the same time.
Several equations can be written down, according to the different
symmetries we want to preserve. A large class of self-consistent
equations is given by the following equation
\begin{equation}\label{4.1}
\langle \psi (i)\psi ^\dag (i)\rangle ={1 \over N}\sum_\mathbf{k} {1
\over {2\pi }}\int_{-\infty }^{+\infty }  d\omega \,C_{\psi \psi
^\dag }(\mathbf{k},\omega )
\end{equation}
where the l.h.s. is fixed by the AC, the WT and the boundary
conditions compatible with the phase under investigation and in the
r.h.s. the correlation function $C_{\psi \psi ^\dag
}(\mathbf{k},\omega )$ is computed by means of the equation of
motion, as illustrated in Section 3. Equations (\ref{4.1}) generate
a set of self-consistent equations which determine the unknown
parameters (i.e., ZFC and unknown correlators) and, consequently,
the proper representation \cite{Mancini_00,Mancini_04}, avoiding the
problem of uncontrolled and uncontrollable decoupling.

\section{Approximation schemes}

The generalized Dyson equation (\ref{2.26}) is an exact equation and
permits, in principle, once the normalization matrix
$I(\mathbf{i},\mathbf{j})$ [cfr. Eq. (2.13)], the m-matrix
$m(\mathbf{i},\mathbf{j})$ [cfr. Eq. (2.14)] and the propagator
$B(i,j)$ [cfr. Eq. (2.24)] are known, in the framework of the
self-consistent scheme outlined in Sections 3 and 4, the calculation
of the various Green's functions. However, for most of the physical
systems of interest the calculation of the propagator $B(i,j)$ is a
very difficult task and some approximations are needed. Various
approximate schemes have been proposed. We will summarize some of
them.

\subsection{The n-pole approximation}

The simplest approximation is based on completely neglecting the
dynamical part $\Sigma (\mathbf{k},\omega )$ [cfr. (2.28)] of the
self-energy. This approximation is largely used in the literature
\cite{Hubbard_63,Hubbard_64,Hubbard_64a,Becker_90,Mori_65,Mori_65a,Fedro_92,Fulde_95,Plakida_89,Mehlig_95,Rowe_68,Roth_69,Beenen_95,Kalashnikov_73,Nolting_72,%
Geipel_88,Nolting_89,Lonke_71,Tserkovnikov_81,Tserkovnikov_81a,Ishihara_94,Mancini_94,Mancini_95,Mancini_95a,Mancini_95b,Avella_98,Shimahara_91,Kruger_94,%
Kondo_72,Yamaji_73} and is called pole-approximation. In this
approximation we only need the knowledge of the normalization matrix
and the $m$-matrix. The constraint (\ref{2.9}) produces a physics of
the solution totally extraneous to the complementary physical space,
orthogonal to that spanned by the multiplet $\psi(i)$. This
approximation, or assumption to a larger extent, consists in
retaining that one can neglect finite life-time effects (i.e., the
dynamical part of the self-energy) paying attention to the choice of
a proper extended operatorial basis, with respect to which the
self-energy corrections have a small total weight. Indeed, the total
weight of the corrections is bounded by the thermal average
(\ref{2.24}) involving the residual source $\delta J (i)$. It is
worth noting \cite{Mancini_98b} that the $n$-pole structure of the
various GF corresponds to a Dyson-like equation
\begin{equation}\label{5.1}
G_{ab}^Q(\mathbf{k},\omega )={{I_{ab}(\mathbf{k})} \over {\omega
-\Sigma _{ab}^Q(\mathbf{k},\omega )}}
\end{equation}
where the self-energy components $\Sigma _{ab}^Q(\mathbf{k},\omega
)$ have a $(n-1)$-pole structure.

\subsection{Self-consistent Born approximation}

In order to improve the approximation one needs to take into account
self-energy corrections by developing some methods to calculate the
effects of $\Sigma (\mathbf{k},\omega )$. In the self-consistent
Born approximation (SCBA), or non-crossing approximation, the
many-particle Green's functions, appearing in the expression of
$\Sigma (\mathbf{k},\omega )$ [see (\ref{2.28})], are calculated by
assuming that the fermionic and bosonic modes propagate
independently.

By recalling the results given in Section 2, the knowledge of the
self-energy requires the calculation of the higher-order propagator
$B^Q(i,j)=\langle Q[\delta J(i)\delta J^\dag (j)]\rangle $. In order
to illustrate the approximation, let us consider the case where the
basic set $\{ \psi (i)\} $ is of a fermionic type. Then, typically
we have to calculate GF of the form $H^R(i,j)=\langle
R[B(i)F(i)F^\dag (j)B^\dag (j)]\rangle $ where $F(i)$ and $B(i)$ are
fermionic and bosonic field operators, respectively. By means of the
spectral representation (\ref{3.20a}) we can write
\begin{equation}\label{5.3}
H^R(\mathbf{k},\omega )=-{1 \over \pi }\int_{-\infty }^{+\infty }
d\omega' {1 \over {\omega -\omega +\mathrm{i}\varepsilon }}\coth
{{\beta \omega' } \over 2}\Im [H^c(\mathbf{k},\omega' )]
\end{equation}
where $H^C(i,j)=\langle T[B(i)F(i)F^\dag (j)B^\dag (j)]\rangle $ is
the causal function. In the SCBA we approximate $H^C(i,j) \approx
f^C(i,j)b^C(i,j)$ where
\begin{equation}\label{5.5}
f^C(i,j)=\langle T[F(i)F^\dag (j)]\rangle \quad \quad \quad
b^C(i,j)=\langle T[B(i)B^\dag (j)]\rangle
\end{equation}

This approximation has been used in many works (for instance see
\cite{Plakida_99,Plakida_01,Avella_03c,Krivenko_04}. By assuming
that the system is ergodic we can use the spectral representations
(\ref{3.21b}) and (\ref{3.38}) to obtain
\begin{align}\label{5.6}
&f^C(\mathbf{k},\omega )=-{1 \over \pi }\int_{-\infty }^{+\infty }
d\omega' [{{1-f_\mathrm{F}(\beta \omega' )} \over {\omega -\omega'
+\mathrm{i}\delta }}+{{f_\mathrm{F}(\beta \omega' )} \over {\omega
-\omega' -\mathrm{i}\delta }}]\Im [f^R(\mathbf{k},\omega )]\\
&b^C(\mathbf{k},\omega )=-{1 \over \pi }\int_{-\infty }^{+\infty }
d\omega' [{{1+f_\mathrm{B}(\beta \omega' )} \over {\omega -\omega
+\mathrm{i}\delta }}-{{f_\mathrm{B}(\beta \omega' )} \over {\omega
-\omega' -\mathrm{i}\delta }}]\Im [b^R(\mathbf{k},\omega )]
\end{align}

Use of (\ref{5.3})-(\ref{5.6}) leads to
\begin{align}\label{5.7}
&H^R(\mathbf{k},\omega )={1 \over \pi }\int_{-\infty }^{+\infty }
d\omega' {1 \over {\omega -\omega' +\mathrm{i}\delta }}{{a^d} \over
{(2\pi )^{d+1}}}\int_{\Omega _B}  d^dp d\Omega \Im
[f^R(\mathbf{p},\Omega
)]\nonumber\\
&\Im [b^R(\mathbf{k}-\mathbf{p},\omega' -\Omega )][\tanh {{\beta
\Omega } \over 2}+\coth {{\beta (\omega' -\Omega )} \over 2}]
\end{align}
where $d$ is the dimensionality of the system, $a$ is the lattice
constant and $\Omega_\mathrm{B}$ is the volume of the Brillouin
zone.

It should be noted that the SCBA can also be applied to the
correlation function. We start from the expression
\begin{equation}\label{5.8}
H^R(\mathbf{k},\omega )={1 \over {2\pi }}\int_{-\infty }^{+\infty }
d\omega' {{1+e^{-\beta \omega' }} \over {\omega -\omega'
+\mathrm{i}\varepsilon }}H(\mathbf{k},\omega' )
\end{equation}
where $H(i,j)=\langle B(i)F(i)F^\dag (j)B^\dag (j)\rangle $ is the
correlation function. In the SCBA we approximate
\begin{equation}\label{5.9}
H(i,j)=\langle B(i)F(i)F^\dag (j)B^\dag (j)\rangle \approx \langle
F(i)F^\dag (j)\rangle \langle B(i)B^\dag (j)\rangle
\end{equation}
Then, by proceeding in the similar way we arrive to the same
expression (\ref{5.7}).

\subsection{Two-site resolvent approach}

In this subsection we will consider an approximation scheme
\cite{Matsumoto_96,Matsumoto_97}, where the dynamical part $\Sigma
(\mathbf{k},\omega )$ [cfr.~(\ref{2.28})] of the self-energy is
estimated by a two-site approximation in combined use with the
resolvent method \cite{Kuramoto_83}. In this approximation the
higher order propagator (\ref{2.24}) is written as
\begin{equation}\label{5.10}
B^Q(\mathbf{k},\omega )=F.T.\langle Q[\delta J(i)\delta J^\dag
(j)]\rangle \approx B_0^Q(\omega )+\alpha (\mathbf{k})B_1^Q(\omega )
\end{equation}
where $B_0^Q(\omega )$ is related to level transitions on equal site
\begin{equation}\label{5.11}
B_0^Q(\omega )={1 \over {2d}}F.T.\langle R[\delta
J(\mathbf{i},t_i)\delta J^\dag (\mathbf{i},t_j)]\rangle
\end{equation}
while $B_1^Q(\omega )$ is related to transitions across two sites
\begin{equation}\label{5.12}
B_1^Q(\omega )={1 \over {2d}}F.T.\langle R[\delta
J(\mathbf{i},t_i)\delta J^{\dag\alpha} (\mathbf{i},t_j)]\rangle
\end{equation}

The Green's function (\ref{2.27}) takes the form
\begin{equation}\label{5.13}
G^Q(\mathbf{k},\omega )={1 \over {\omega -\varepsilon
(\mathbf{k})+t^2V(\omega )\alpha (\mathbf{k})}}I(\mathbf{k})
\end{equation}
where $V(\omega )$ has to be calculated from the definition
(\ref{2.28}) by making use of approximation (\ref{5.10}). We give
now a brief sketch of the calculation.

$B_0^Q(\omega )$ and $B_1^Q(\omega )$ are computed by expressing
$\delta J(i)$ in terms of transitions among the two-site levels
\begin{equation}\label{5.14}
\delta J=\sum_{nm}  a_{nm}\Phi _n^\dag \Phi _m
\end{equation}
where $\{ \Phi _m\} $ is the complete set of operators for two-site
levels. By means of the non crossing approximation
\cite{Kuramoto_83,Grewe_83}, the propagator $\langle Q[\Phi _n^\dag
(t_i)\Phi _m(t_i)\Phi _{n}^\dag (t_j)\Phi _m(t_j)]\rangle $ is
expressed in terms of the resolvent $R_{nm}(t_i-t_j)=\langle Q[\Phi
_n(t_i)\Phi _m^\dag (t_j)]\rangle _R$ where  the subscript $R$
indicates the reservoir system, i.e., the part $H_R$ of the
Hamiltonian other than that concerned with two sites. The
calculation of the resolvent brings to a modification of the
original two-site levels by the surroundings. In this scheme effects
of time delay in the local correlations are treated trough the
time-dependent modifications of the two-site level transitions,
which are included as time-dependent local effects in the electron
self-energy.

\section{Conclusions}

In this article we have illustrated an approach to the study of
highly correlated electronic systems, based on the equation of
motion and Green's function method.  Such an approach is based on
two main ideas: (i) propagators of composite operators as building
blocks at the basis of approximate calculations; (ii) use of algebra
constrains to fix the representation of the GF in order to maintain
the algebraic and symmetry properties. This formalism has been
applied to the study of several models of highly interacting
systems, and we refer the interested readers to
Ref.~\cite{Mancini_04} for an exhaustive bibliography.


\label{last@page}
\end{document}